\documentclass[preprint,showpacs,preprintnumbers,amsmath,amssymb]{revtex4}
\usepackage{amsmath}

\usepackage{graphicx}
\usepackage{dcolumn}
\usepackage{bm}
\usepackage{amsmath}

\newcommand{\qed}{\nobreak \ifvmode \relax \else
      \ifdim\lastskip<1.5em \hskip-\lastskip
      \hskip1.5em plus0em minus0.5em \fi \nobreak
      \vrule height0.75em width0.5em depth0.25em\fi}

\begin{document}

\preprint{}

\title{A Balanced $_{10}F_{9}$ Hypergeometric Hybrid Hilbert-Schmidt/Bures Two-Qubit Function and Related Constructions}
\author{Paul B. Slater}%
\email{slater@kitp.ucsb.edu}
\affiliation{%
University of California, Santa Barbara, CA 93106-4030\\
}%
\date{\today}

\begin{abstract}
We seek to develop a Bures (minimal monotone/statistical distinguishability) metric-based series of formulas for the moments of probability distributions over the determinants $|\rho|$ and $|\rho^{PT}|$ of $4 
\times 4$ density matrices, $\rho$, for generalized 
(rebit, quater[nionic]bit,\ldots) two-qubit systems, analogous to a series that  has been obtained for the Hilbert-Schmidt (HS) metric.  In particular, we desire--using moment-inversion procedures--to be able to closely test the previously-developed conjecture ({\it J. Geom. Phys.}, {{\bf{53}}}, 74 [2005]) that the Bures separability probability over the (standard, fifteen-dimensional convex set of) two-qubit states is $\frac{1680 \left(\sqrt{2}-1\right)}{\pi ^8} \approx 0.0733389$--while, in the HS context, strong evidence has been adduced, along the indicated analytical lines, that the counterpart of this value is $\frac{8}{33}$ ({\it J. Phys. A}, {{\bf45}}, 095305 [2012]). Working within the "utility function" framework of Dunkl 
employed in that latter study, we obtain
an interesting $_{10}F_{9}$ balanced hypergeometric function based on a "hybridization" of 
known Bures and HS terms. This exercise appears to provide an upper bound on the Bures two-qubit separability probability of 0.0798218. We also examine the yet unresolved HS qubit-qutrit scenario. Mathematica calculations indicate that if the same form of hypergeometric paradigm as has been established for the generalized two-qubit HS moments is followed in either the HS 
qubit-qutrit or Bures two-qubit cases, then a balanced hypergeometric function 
$_{p}F_{p-1}$ with $p>9$ would be required.
\end{abstract}


\maketitle
\section{Introduction}
We seek to expand upon an extended line of work that has been reported in 
\cite{MomentBased,BuresHilbert,slaterJModPhys,SlaterFisher}.
Our immediate motivation is to find, if possible, Bures-metric-based counterparts to results 
(still not all rigorously demonstrated) that have been developed in terms of the Hilbert-Schmidt metric \cite{szBures,szHS,ingemarkarol}. These assertions concern Hilbert-Schmidt probability distributions over the determinants $|\rho|$ and $|\rho^{PT}|$ 
(and their joint products) of $4 \times 4$ density matrices 
$\rho$ for generalized (rebit, quater(nionic)bit,\ldots) two-qubit states, where $PT$ denotes the partial transpose. The approach that has been adopted involves, first,  finding--with the aid of symbolic computations--the moments of these  probability distributions, and  second, using Legendre-polynomial-based moment-inversion procedures \cite{Provost} to generate the associated probabilities that a state is separable (that is, $|\rho^{PT}|>0$) \cite{ZHSL,Simon}, as well as the probability densities at the entanglement-separability boundary (the locus of 
$|\rho^{PT}|=0$) \cite{SlaterFisher}. Explicit formulas have not so far been developed for the underlying Hilbert-Schmidt probability distributions themselves--as opposed to these two interesting specific features.

To begin, we note corrigenda for our previous report \cite{BuresHilbert} to the following two companion  formulas there (eqs. (8) and (25)),
\begin{equation} \label{explanatory}
\frac{\left\langle \left\vert \rho^{PT}\right\vert \left\vert \rho\right\vert
^{k}\right\rangle^{Bures}_{2-qubits}}{\left\langle \left\vert \rho\right\vert ^{k}\right\rangle^{Bures}_{2-qubits}}
=\frac{8 k^5+36 k^4-82 k^3-681 k^2-1366 k-885}{128 \left(16 k^5+192 k^4+883 k^3+1947
   k^2+2062 k+840\right)}
\end{equation}
and
\begin{equation} \label{RebitExplanatory}
\frac{\left\langle \left\vert \rho^{PT}\right\vert \left\vert \rho\right\vert
^{k}\right\rangle^{Bures}_{2-rebits}}{\left\langle \left\vert \rho\right\vert ^{k}\right\rangle^{Bures}_{2-rebits}}
= \frac{64 k^5+128 k^4-340 k^3-1032 k^2-1099 k-384}{k \left(8 k^2-2
   k-1\right) \left(8 k^2+18 k-5\right)},
\end{equation}
where the brackets indicate expected values (averages) over the corresponding 15-dimensional and 9-dimensional convex sets of states, respectively.
This pair of formulas had been constructed with the aid of the FindSequenceFunction command of Mathematica, with the first number of each of the two sequences being inputted,  corresponding to the case $k=0$ \cite[eq. (2)]{BuresHilbert}. However, the command (naturally enough) took these numbers to correspond to the case $k=1$. The first formula must be modified--as we have subsequently observed--by replacing $k$ in it by $k+1$. Such a transformation yields 
\begin{equation} \label{explanatory2}
\frac{\left\langle \left\vert \rho^{PT}\right\vert \left\vert \rho\right\vert
^{k}\right\rangle^{Bures}_{2-qubits}}{\left\langle \left\vert \rho\right\vert ^{k}\right\rangle^{Bures}_{2-qubits}}
= \frac{8 k^5+76 k^4+142 k^3-631 k^2-2790 k-2970}{2048 k^5+34816 k^4+231808
   k^3+756224 k^2+1209984 k+760320} =
\end{equation}
\begin{displaymath}
\frac{k \left(k \left(2 k \left(4 k^2+38
   k+71\right)-631\right)-2790\right)-2970}{128 (k+3) (k+4) (k+5) (4 k+9) (4
   k+11)}.
\end{displaymath}
The second formula must be modified by this same transformation, followed by multiplication by $\frac{\left\langle \left\vert \rho\right\vert ^{k+1}\right\rangle^{Bures}_{2-rebits}}{\left\langle \left\vert \rho\right\vert ^{k}\right\rangle^{Bures}_{2-rebits}}$, where
\begin{equation} \label{BuresAverage}
\left\langle \left\vert \rho\right\vert ^{k}\right\rangle^{Bures}_{\alpha} =\frac{2^{-4 \alpha-8 k-1} \Gamma (6 \alpha+2) \Gamma \left(k+\frac{1}{2}\right) \Gamma
   \left(\alpha+k+\frac{1}{2}\right) \Gamma (\alpha+2 k+1)}{\sqrt{\pi } \Gamma (2 \alpha+k+1)
   \Gamma (3 \alpha+k+1) \Gamma \left(3 \alpha+2 k+\frac{3}{2}\right)}=
\end{equation}
\begin{displaymath}
\frac{\left(\frac{1}{2}\right)_k 
\left(\alpha +\frac{1}{2}\right)_k
   (\alpha +1)_{2 k}}{2^{8 k} (2 \alpha +1)_k (3 \alpha +1)_k \left(3 \alpha
   +\frac{3}{2}\right)_{2 k}},
\end{displaymath}
with the Dyson-index-like parameter $\alpha$ set to  = $\frac{1}{2}$ for the two-rebit scenario.
Taking these steps yields
\begin{equation} \label{RebitExplanatory2}
\frac{\left\langle \left\vert \rho^{PT}\right\vert \left\vert \rho\right\vert
^{k}\right\rangle^{Bures}_{2-rebits}}{\left\langle \left\vert \rho\right\vert ^{k}\right\rangle^{Bures}_{2-rebits}}
= \frac{64 k^5+448 k^4+812 k^3-644 k^2-3351 k-2663}{16384 k^5+188416 k^4+847872
   k^3+1869824 k^2+2023424 k+860160} = 
\end{equation}
\begin{displaymath}
\frac{k (4 k (k (16 k (k+7)+203)-161)-3351)-2663}{2048 (k+2)^2 (2 k+3) (2 k+5)
   (2 k+7)}.
\end{displaymath}
(Further, let us indicate that the numerator factor of $32^{2 -8 k}$ in equation (26) in \cite{BuresHilbert} should have been more clearly 
written as $3 \cdot 2^{2 -8 k}$. Still further, we note that the final term $(\alpha +k+2)_{n-j}$ in the second equation on p. 30 in \cite{MomentBased} should be  
$(2 \alpha +k+1)_{n-j}$.)

Now, for $k=0$, we obtain, quite interestingly, the averages (first moments) $\left\langle \left\vert \rho^{PT}\right\vert \right\rangle^{Bures}_{2-qubits}=
-\frac{1}{256}=-2^{-8}= -\frac{2970}{760320}=\frac{8}{2048}$ (that is, the common 
ratio of the zeroth-degree, as well as the fifth-degree coefficients in 
(\ref{explanatory2})) and  $\left\langle \left\vert \rho^{PT}\right\vert \right\rangle^{Bures}_{2-rebits}=
-\frac{2663}{860160}$ (the ratio of the constant terms in (\ref{RebitExplanatory2})), as had been reported in \cite[eqs. (2), (18)]{BuresHilbert}. The ratio of the fifth-degree coefficients in the two-rebit case is, once again,
$\frac{1}{256}=\frac{64}{16384}$.

The corrected formula (\ref{explanatory2}) now fully reproduces the exact moment calculations for $\left\langle  \left\vert  \rho^{PT}\right\vert  \left\vert  \rho \right\vert^{k}  \right\rangle_{2-qubits}^{Bures}$, $k=0,\ldots,5$, explicitly reported previously \cite[eqs. (2)-(7)]{BuresHilbert}. Similarly, the corrected formula 
(\ref{RebitExplanatory2}) fully reproduces the exact moment calculations for 
$\left\langle  \left\vert  \rho^{PT}\right\vert  \left\vert  \rho \right\vert^{k}  \right\rangle_{2-rebits}^{Bures}$, $k=0,\ldots 4$, also explicitly reported \cite[eqs. (20)-(24)]{BuresHilbert}.

Two of our first findings below (secs.~\ref{Hyb1} and \ref{Hyb2})--in regard to ${10}F_{9}$ hypergeometric functions--are independent of the (amended) moment formulas above, depending only upon the "utility function" algebraic framework--developed by 
Dunkl--in the earlier generalized
Hilbert-Schmidt-two-qubit study \cite[sec. D.4]{MomentBased}. In sec.~\ref{8F12} we do report the construction of an $_{8}F_{12}$-based function that does succeed (but in a perhaps somewhat ad hoc fashion) in jointly reproducing these two formulas. Previous to this, in sec.~\ref{Trial}, we construct a $_{5}F_{4}$-based expression that closely, but not fully, reproduces the two target formulas.

In one sense, our objectives here might be regarded as "overly-ambitious". That is, in contrast to the Hilbert-Schmidt study \cite{MomentBased}--we have, in view of the substantial Bures computational (high-precision numerical integration,\ldots) challenges, quite limited determinantal moment calculations at hand \cite{BuresHilbert} to use for modeling purposes. (We do report, however, our apparently successful attempt to compute 
$\frac{\left\langle \left\vert \rho^{PT}\right\vert^2 \left\vert \rho\right\vert
^{k}\right\rangle^{Bures}_{2-rebits}}{\left\langle \left\vert \rho\right\vert ^{k}\right\rangle^{Bures}_{2-rebits}}$ \cite[eqs. (33)-(37)]{BuresHilbert} [sec.~\ref{Arduous}].) So, we may lack sufficient information to reach truly definitive conclusions. But, in light of the intrinsic interest in the separability probability question \cite{ZHSL}, we will try to maximize our insights and potential advances, given the presently available results. (We hope eventually, however, to be able to expand the inventory of explicit Bures determinantal moment calculations, and further utilize them in Bures two-qubit separability probability estimations.)
\section{Analyses}
\subsection{Bures Measure}
With these initial remarks, we now seek to find the Bures (minimal monotone metric) counterpart to the centrally-important Hilbert-Schmidt (HS) moment formula 
\cite[App. D.6]{MomentBased} (cf. \cite[eq. (17)]{Dehesa}) ("any ball with respect to the Bures distance of a fixed radius in the space of quantum states has the same measure" \cite{osipov}), 
\begin{equation}  \label{BigOne}
\frac{\left\langle \left\vert \rho^{PT}\right\vert ^{n}\left\vert \rho\right\vert
^{k}\right\rangle^{HS}_{\alpha}}{\left\langle \left\vert \rho\right\vert ^{k}\right\rangle^{HS}_{\alpha}}
  =\frac{\left(  k+1\right)  _{n}\left(  k+1+\alpha\right)  _{n}\left(
k+1+2\alpha\right)  _{n}}{2^{6n}\left(  k+3\alpha+\frac{3}{2}\right)
_{n}\left(  2k+6\alpha+\frac{5}{2}\right)  _{2n}}\\
\end{equation}
\begin{displaymath}
 \times~_{5}F_{4}\left(
\genfrac{}{}{0pt}{}{-n,-k,\alpha,\alpha+\frac{1}{2},-2k-2n-1-5\alpha
}{-k-n-\alpha,-k-n-2\alpha,-\frac{k+n}{2},-\frac{k+n-1}{2}}%
;1\right) , 
\end{displaymath}
where
\begin{equation} \label{generalformula}
\left\langle \left\vert \rho\right\vert ^{k}\right\rangle^{HS}_{\alpha} =\frac{k!\left(
\alpha+1\right)  _{k}\left(  2\alpha+1\right)  _{k}}{2^{6k}\left(
3\alpha+\frac{3}{2}\right)  _{k}\left(  6\alpha+\frac{5}{2}\right)  _{2k}},
\end{equation}
is the Hilbert-Schmidt counterpart of the Bures formula (\ref{BuresAverage}), and the Pochhammer symbol notation is used.
(The Dyson-index-like parameter $\alpha$ \cite{MatrixModels} assumes the value $\frac{1}{2}$ for the nine-dimensional convex set of two-rebit systems, the value 1
for the [standard] fifteen-dimensional convex set of two-qubit systems, and 2 in the twenty-seven-dimensional 
 quater(nionic)bit instance,\ldots. It is certainly a yet formally unresolved question whether or not the Dyson-index ansatz does extend from the HS to the Bures framework.)

We can, now, deduce that for the classical ($\alpha=0$) cases (where $|\rho^{PT}|=|\rho|$, with $\rho$ being simply diagonal in nature) that the general formulas of interest reduce there to (with $n=1$)
\begin{equation} \label{HSclassical}
\frac{\left\langle \left\vert \rho \right\vert \left\vert \rho\right\vert
^{k}\right\rangle^{HS}_{\alpha=0}}{\left\langle \left\vert \rho\right\vert ^{k}\right\rangle^{HS}_{\alpha=0}} = \frac{k^3+3 k^2+3 k+1}{256 k^3+1152 k^2+1712 k+840} = \frac{(k+1)^3}{8 (2 k+3) (4 k+5) (4 k+7)}
\end{equation}
and 
\begin{equation} \label{Buresclassical}
\frac{\left\langle \left\vert \rho \right\vert \left\vert \rho\right\vert
^{k}\right\rangle^{Bures}_{\alpha=0}}{\left\langle \left\vert \rho\right\vert ^{k}\right\rangle^{Bures}_{\alpha=0}} = \frac{8 k^3+12 k^2+6 k+1}{2048 k^3+6144 k^2+6016 k+1920} = \frac{(2 k+1)^3}{128 (k+1) (4 k+3) (4 k+5)}.
\end{equation}
\subsection{Trial expression} \label{Trial}
At an early state in our  efforts  to reproduce formulas (\ref{explanatory2}) and (\ref{RebitExplanatory2})--emulating the framework employed in  \cite{MomentBased}--we used for 
$\frac{\left\langle \left\vert \rho^{PT}\right\vert ^{n}\left\vert \rho\right\vert
^{k}\right\rangle^{Bures}_{\alpha}}{\left\langle \left\vert \rho\right\vert ^{k}\right\rangle^{Bures}_{\alpha}}$ the expression
\begin{equation} \label{BigOne2}
 \frac{\left(k+\frac{1}{2}\right)_n \left(k+\alpha +\frac{1}{2}\right)_n
   (2 k+\alpha +1)_{2 n}}{2^{8 n}(k+2 \alpha +1)_n (k+3 \alpha +1)_n \left(2 k+3 \alpha
   +\frac{3}{2}\right)_{2 n}}\\
\end{equation}
\begin{displaymath}
  \times~_{6}F_{5}\left(
\genfrac{}{}{0pt}{}{-n,-k,\alpha,\alpha+\frac{1}{2},-2k-2n-4-8\alpha,-2k-2n -3 -2 \alpha}{-k-n,-k-n-4\alpha,-\frac{k+n}{2}+\frac{1}{4},-\frac{k+n}{2}+\frac{1}{2},-2k-2n -5 }%
;1\right)  .
\end{displaymath}
(We note that this hypergeometric term  is not fully "balanced" [in the first, predominant "Saalschutzian" sense of this term \cite[p. 6610]{Dehesa}], as its counterpart (\ref{BigOne}) is in the Hilbert-Schmidt case 
\cite[sec. D.4]{MomentBased}.)
This formulation gives us--inserting $n=1$, and $\alpha=1$ 
(two-qubit) and $\alpha = \frac{1}{2}$ (two-rebit), respectively--
\begin{equation} \label{fit1}
\frac{8 k^5+76 k^4+142 k^3-631 k^2-2028 k-1467}{2048 k^5+34816 k^4+231808
   k^3+756224 k^2+1209984 k+760320}
\end{equation}
and
\begin{equation} \label{fit2}
\frac{64 k^5+448 k^4+812 k^3-644 k^2-2231 k-1095}{16384 k^5+188416 k^4+847872
   k^3+1869824 k^2+2023424 k+860160}.
\end{equation}
So, we see that the target formulas (\ref{explanatory2}) and (\ref{RebitExplanatory2}) above are reproduced, but for the two lowest-degree terms ([$-2028 k-1467$] and [$-2231 k-1095$] {\it vs.} [$-2790 k -2970$] and [$-3351 k -2663$]) in the two numerators. Also, for 
$\alpha=0$ the classical formula
(\ref{Buresclassical}) is fully reproduced by this trial hypergeometric-based expression (\ref{BigOne2}).

Of course, our natural objective is to be able to successfully reproduce
these pairs of lowest-degree numerator terms in  
(\ref{explanatory2}) and (\ref{RebitExplanatory2}), as well (cf. sec.~\ref{8F12}). (We speculate that, for general $n$--only the case $n=1$ is discussed above, being the single value of $n$ for which exact computations have been fully performable \cite{BuresHilbert} for 
$\alpha=\frac{1}{2}$ and 1--the presumed hypergeometric-based formula, in the Bures case, will yield "biproper" rational functions \cite{chou} being composed of degree-($4 n+1$) numerator and denominator polynomials. In the Hilbert-Schmidt generalized two-qubit case \cite{MomentBased}, the corresponding numerator and denominator 
polynomials of the rational functions are both of 
degree-$3 n$, while they--based on the $n=1$ and $n=2$ results reported in  \cite[sec. 6]{MomentBased}--appear to be of 
degree-($4 n+1$) in  the $6 \times 6$ qubit-qutrit density matrix setting.).

Let us insert our newly-developed ("partially successful") hypergeometric-based formula 
(\ref{BigOne2}) (converted to "balanced" [in a second, less common non-Saalschutzian sense of the term] form \cite[sec. 3]{MomentBased}, meaning that we set $n=k$) into the moment-inversion procedure 
\cite{Provost} we have previously extensively used \cite{MomentBased,BuresHilbert,slaterJModPhys,SlaterFisher}. (We note that the range of
the associated variable $|\rho^{PT}| |\rho|$ is $[-2^{-12} 3^{-3},2^{-16}]$.)
Incorporating 3,000 such presumptive "moments", we obtain estimates for the Bures-based separability probabilities of 0.39593 for the two-rebit case ($\alpha=\frac{1}{2}$) and 0.25861 for the two-qubit ($\alpha=1$) case. With 6,000 "moments", these figures diminish somewhat to  0.38718 and 0.23634, respectively.
(In the previously-studied Hilbert-Schmidt instance, the evidence is quite strong that these two probabilities are $\frac{29}{64} \approx 0.453125$ and $\frac{8}{33} \approx 0.242424$, respectively \cite{slaterJModPhys}. It certainly appears, since the Bures measure is more concentrated on states of higher purity \cite[sec. 14.6]{MomentBased}, that the Bures versions of  these rational probabilities should, in fact, be smaller.)

In fact, in \cite[eq. (16), Table VI]{slaterJGP},  it had been hypothesized--based upon extensive 
numerical analyses--that the Bures two-qubit separability probability is 
$\frac{1680 \left(\sqrt{2}-1\right)}{\pi ^8} \approx 0.0733389$. It, then, remained  to see  if we could develop a more successful formula than (\ref{BigOne2}) for $\frac{\left\langle \left\vert \rho^{PT}\right\vert ^{n}\left\vert \rho\right\vert
^{k}\right\rangle^{Bures}_{\alpha}}{\left\langle \left\vert \rho\right\vert ^{k}\right\rangle^{Bures}_{\alpha}}$ that yielded the two lowest-degree numerator terms in (\ref{explanatory2}) and (\ref{RebitExplanatory2}) (cf. sec.~\ref{8F12}), as well as the other higher-degree terms, whether a more satisfactory agreement than 0.23634 {\it vs.} 0.073389, would then ensue. (If we had employed $|\rho^{PT}| \in [-2^{-4},2^{-8}]$ as our principal variable, there are strong indications both from \cite{MomentBased} and sec.~\ref{SepProbEst}
 that the convergence would have 
been considerably more rapid, and to considerably smaller probabilities.)
\subsection{Reformulated goal}
Our major analytic objective has been to ascertain a hypergeometric function $_{p}F_{p-1}$ (where $p$ is yet undetermined--while $p=5$ had been found in the Hilbert-Schmidt case \cite{MomentBased}), parallel to (\ref{BigOne}), that will fully reproduce the Bures two-qubit and two-rebit formulas (\ref{explanatory2}) and 
(\ref{RebitExplanatory2}). An equivalent--but perhaps more computationally direct--approach is to attempt to reproduce these two formulas after
the (non-hypergeometric) prefactor 
\begin{equation} \label{prefactorQQ}
\frac{\left(k+\frac{1}{2}\right)_n \left(k+\alpha +\frac{1}{2}\right)_n
   (2 k+\alpha +1)_{2 n}}{2^{8 n}(k+2 \alpha +1)_n (k+3 \alpha +1)_n \left(2 k+3 \alpha
   +\frac{3}{2}\right)_{2 n}}
\end{equation}
in (\ref{BigOne2}) has been factored out.
Then, the formulas that we aim to reproduce (\ref{explanatory2}) and (\ref{RebitExplanatory2}) become  (in the two-qubit, $\alpha=1$ case),
\begin{equation} \label{simplerqubit}
\frac{8 k^5+76 k^4+142 k^3-631 k^2-2790 k-2970}{8 k^5+76 k^4+238 k^3+329 k^2+204
   k+45}=
\end{equation}
\begin{displaymath}
\frac{k \left(k \left(2 k \left(4 k^2+38
   k+71\right)-631\right)-2790\right)-2970}{(k+1) (k+5) (2 k+1) (2 k+3)^2}
\end{displaymath}
and (in the two-rebit, $\alpha =\frac{1}{2}$ case),
\begin{equation} \label{simplerrebit}
\frac{64 k^5+448 k^4+812 k^3-644 k^2-3351 k-2663}{64 k^5+448 k^4+1068 k^3+1148
   k^2+569 k+105} =
\end{equation}
\begin{displaymath}
-\frac{136}{2 k+1}-\frac{8}{495 (2 k+7)}+\frac{448}{9 (4 k+5)}+\frac{5824}{44
   k+33}-\frac{384}{5 (k+1)}+1,
\end{displaymath}
with the two highest-degree numerator and denominator coefficients now matching in both cases.
C. Dunkl observed that the two immediately preceding expressions (\ref{simplerqubit}) and (\ref{simplerrebit}) could jointly be expressed in terms of the single expression
\begin{equation} \label{Joint}
J(\alpha) =
-\frac{\alpha  (2 \alpha +1) (4 \alpha -1) (4 \alpha +1)}{(k+1) (2 k+1) (\alpha
   +2 k+2) (3 \alpha +k+2) (3 \alpha +2 k)}
\end{equation}
\begin{displaymath}
-\frac{2 \alpha  (2 \alpha +1) (4
   \alpha +4 k+5) (8 \alpha +4 k+3)}{(k+1) (2 k+1) (\alpha +2 k+2) (3 \alpha +2
   k)}+1.
\end{displaymath} 
Dunkl did subsequently point out that this  formula was deficient in its properties
$\alpha \rightarrow \infty$, in the sense that the number of occurrences of $\alpha$'s in its (collected) numerator and denominator do not cancel--as they successfully do in (\ref{BigOne})--the number of occurrences of $\alpha$'s in the prefactor (\ref{prefactorQQ}) in  (\ref{BigOne2}).
\subsection{Utility functions}
We have, further, investigated the possibility of developing Bures analogues of the
Hilbert-Schmidt-based "utility functions" that were of central importance in the predecessor study in uncovering the  Hilbert-Schmidt moment formulas \cite[sec. D.4]{MomentBased}. We incorporate the notational/algebraic framework  established there, but now decorating with Bures superscripts (where HS superscripts were implicitly understood previously in \cite{MomentBased}),
\begin{align*}
F_{0}^{Bures}\left(  k\right)   &  =\left\langle \left\vert
\rho\right\vert ^{k}\right\rangle^{Bures}_{\alpha}=\frac{\left(\frac{1}{2}\right)_k 
\left(\alpha +\frac{1}{2}\right)_k
   (\alpha +1)_{2 k}}{2^{8 k} (2 \alpha +1)_k (3 \alpha +1)_k \left(3 \alpha
   +\frac{3}{2}\right)_{2 k}},\\
F_{1}^{Bures}\left(  n,k\right)   &  =\left\langle \left\vert \rho^{PT}\right\vert
^{n}\left\vert \rho\right\vert ^{k}\right\rangle /\left\langle \left\vert
\rho\right\vert ^{k}\right\rangle ,\\
F_{2}^{Bures}\left(  n,k\right)   &  =\left\langle \left\vert \rho\right\vert
^{k}\left(  \left\vert \rho^{PT}\right\vert -\left\vert \rho\right\vert
\right)  ^{n}\right\rangle /\left\langle \left\vert \rho\right\vert
^{k}\right\rangle ,\\
R^{Bures}\left(  n,k\right)   &  =F_{0}^{Bures}\left(  n+k\right)  /F_{0}^{Bures}\left(  k\right).
\end{align*}
with the same form of interesting decomposition
\begin{equation} \label{Utility1}
\left\vert \rho^{PT}\right\vert ^{n}   =\sum_{j=0}^{n}\binom{n}{j}\left\vert
\rho\right\vert ^{n-j}\left(  \left\vert \rho^{PT}\right\vert -\left\vert
\rho\right\vert \right)  ^{j}
\end{equation}
holding, and
\begin{equation} \label{Summation}
F_{1}^{Bures}\left(  n,k\right)    =\sum_{j=0}^{n}\binom{n}{j}F_{2}^{Bures}\left(
j,k+n-j\right)  R^{Bures}\left(  n-j,k\right),  
\end{equation}
with the basic objective being to compute $F_{1}^{Bures}\left(  n,k\right)  $.
Adding a third argument to these functions of 
$\alpha = 1$ or $\frac{1}{2}$  to denote the two-qubit and two-rebit cases, respectively, we were able to find, using (\ref{explanatory2}) and (\ref{RebitExplanatory2}) for the $F_{1}^{Bures}(1,k,\alpha)$ functions, that 
\begin{equation} \label{F2Complex}
F_2^{Bures}(1,k,1) = \frac{-96 k^3-960 k^2-2994 k-3015}{2048 k^5+34816 k^4+231808 k^3+756224 k^2+1209984
   k+760320}=
\end{equation}
\begin{displaymath}
-\frac{3 (2 k (16 k (k+10)+499)+1005)}{128 (k+3) (k+4) (k+5) (4 k+9) (4 k+11)}.
\end{displaymath}
and
\begin{equation} \label{F2Real}
F_2^{Bures}(1,k,\frac{1}{2}) = \frac{-16 k^3-112 k^2-245 k-173}{1024 k^5+11776 k^4+52992 k^3+116864 k^2+126464
   k+53760} = 
\end{equation}
\begin{displaymath}
-\frac{k (16 k (k+7)+245)+173}{128 (k+2)^2 (2 k+3) (2 k+5) (2 k+7)}.
\end{displaymath}
As in the Hilbert-Schmidt case, these rational functions are of degree -2, that is, the degree of the denominator
is two more than the degree of the numerator.
In the Bures instance, the degrees of both numerators  are three, and the denominator degrees, five. However, in the simpler HS case, the numerators were  of degree zero, and the denominators, degree two. It, thus, appeared that more explicit Bures moment computations ($n>1$) were  needed than available from \cite{BuresHilbert} to effectively pursue this utility function route further. 

Using the formula for $J(\alpha)$ 
given by (\ref{Joint}), we have (similarly encompassing  (\ref{F2Complex}) and 
(\ref{F2Real}) into a single formula) 
\begin{equation}
F_2^{Bures}(1,k,\alpha)= J(\alpha)-\frac{(2 k+1) (\alpha +2 k+1) (\alpha +2 k+2) (2 \alpha +2 k+1)}{256 (2 \alpha
   +k+1) (3 \alpha +k+1) (6 \alpha +4 k+3) (6 \alpha +4 k+5)}.
\end{equation}
This formula for $F_2^{Bures}(1,k,\alpha)$ needs to be extended to one for $F_2^{Bures}(n,k,\alpha)$, so that the summation in (\ref{Summation}) would be fully properly expressed and conducted.

We, interestingly, note (results of the WolframAlpha website) that the numerator of the two-qubit result (\ref{F2Complex}) can be expressed as
\begin{equation}
-96 \left(k+\frac{10}{3}\right)^3+206 \left(k+\frac{10}{3}\right)-\frac{1315}{9},
\end{equation}
while the numerator of the two-rebit result (\ref{F2Real}) can be represented as 
\begin{equation}
-16 \left(k+\frac{7}{3}\right)^3+\frac{49}{3}
   \left(k+\frac{7}{3}\right)-\frac{212}{27}
\end{equation}
or
\begin{equation}
-\frac{1}{4} (4 k+7)^3-\frac{7}{4} (4 k+7)^2-\frac{3}{2}.
\end{equation}
The expression
\begin{equation} \label{DenominatorFunction}
\mbox{den}(\alpha)= 128 \alpha  (2 \alpha +k+1) (3 \alpha +k+1) (3 \alpha +k+2) (6 \alpha +4 k+3)
   (6 \alpha +4 k+5)
\end{equation}
yields, for 
$\alpha=\frac{1}{2}$, the denominator of $F_2^{Bures}(1,k,\frac{1}{2})$, given by (\ref{F2Real}), and for $\alpha=1$, the denominator of $F_2^{Bures}(1,k,1)$, given by (\ref{F2Complex}). Similarly, we have the following expression yielding the numerators
of these two $F_2^{Bures}(1,k,\alpha)$ functions,
\begin{equation} \label{NumeratorFunction}
\mbox{num}(\alpha)= \frac{1}{27} \left(-7466 \alpha -32 (5 \alpha -2) (6 \alpha +3 k+4)^3+6 (569
   \alpha -260) (6 \alpha +3 k+4)+3521\right).
\end{equation}
(It, thus, appears that $\mbox{num}(\alpha)$, unfortunately for further analytical purposes (sec.~\ref{8F12}), lacks as simple a form of factorization as $\mbox{den}(\alpha)$.)
We will employ these last two functions, $\mbox{num}(\alpha)$ and $\mbox{den}(\alpha)$, in an interesting exercise in sec.~\ref{8F12}.
\subsection{Hybrid Hilbert-Schmidt/Bures $_{10}F_{9}$ function} \label{Hyb1}
We have also determined, in the context of the generalized Hilbert-Schmidt two-qubit results of \cite{MomentBased} that
\begin{equation} \label{F2HS}
F_2^{HS}(n,k,\alpha)=\frac{2^{-6 n} (\alpha )_n \left(\alpha +\frac{1}{2}\right)_n (-2 k-2 n-5 \alpha
   -1)_n}{\left(k+3 \alpha +\frac{5}{4}\right)_n \left(k+3 \alpha
   +\frac{3}{2}\right)_n \left(k+3 \alpha +\frac{7}{4}\right)_n}.
\end{equation}
Further, we are immediately able to construct from (\ref{generalformula}) the utility 
function
\begin{equation}
R^{HS}(n,k,\alpha)=
\frac{2^{-6 n} \left(3 a+\frac{3}{2}\right)_k \left(6 a+\frac{5}{2}\right)_{2 k}
   (1)_{k+n} (a+1)_{k+n} (2 a+1)_{k+n}}{(1)_k (a+1)_k (2 a+1)_k \left(3
   a+\frac{3}{2}\right)_{k+n} \left(6 a+\frac{5}{2}\right)_{2 (k+n)}}
\end{equation}
and from (\ref{BuresAverage}) \begin{equation} \label{RBures}
R^{Bures}(n,k,\alpha)=
\frac{2^{-8 n} \left(\frac{1}{2}\right)_{k+n} (2 \alpha +1)_k (3 \alpha +1)_k
   \left(3 \alpha +\frac{3}{2}\right)_{2 k} \left(\alpha
   +\frac{1}{2}\right)_{k+n} (\alpha +1)_{2 (k+n)}}{\left(\frac{1}{2}\right)_k
   \left(\alpha +\frac{1}{2}\right)_k (\alpha +1)_{2 k} (2 \alpha +1)_{k+n} (3
   \alpha +1)_{k+n} \left(3 \alpha +\frac{3}{2}\right)_{2 (k+n)}}.
\end{equation}
Let us perform the principal utility function summation indicated in eq. (\ref{Summation}), previously performed in \cite{MomentBased} with the Hilbert-Schmidt counterparts, but now replacing the (unknown) term $F_2^{Bures}(j,k+n-j)$ by its known Hilbert-Schmidt counterpart (\ref{F2HS}), while retaining $R^{Bures}(n,k,\alpha)$, given by (\ref{RBures}). Then, a rather lengthy computation produced our first main ("hybrid") result reported here. 

This  result was identically equal to the product of the Bures prefactor in (\ref{BigOne2}) times a certain $_{10}F_{9}$ hypergeometric function (with argument 1) that is, remarkably, exactly balanced/Saalschutzian--that is, the sum of the ten numerator terms plus 1 equals the
sum of the nine denominator terms. 
The result also possesses the desired property (for asymptotic [$\alpha \rightarrow \infty$] analytic reasons, as indicated by C. Dunkl) of having the parameter $\alpha$ contained in one more (that is, seven) of the ten numerator terms than it is contained in (that is, six) of the nine denominator terms. (This successfully compensates--paralleling the structure of the earlier Hilbert-Schmidt result (\ref{BigOne}) 
\cite[p. 30]{MomentBased}--for the prefactor in (\ref{BigOne2}) which, oppositely, has one more occurrence of 
$\alpha$--that is, four--in the denominator than numerator, which has three.) 

The ten numerator terms of this $_{10}F_{9}$ function were 
\begin{equation}
\{\alpha ,\alpha +\frac{1}{2},-n,-k,-5 \alpha -2 k-2 n-1,-k-n,-3 \alpha -k-n,-2 \alpha -k-n,
\end{equation}
\begin{displaymath}
-\frac{3 \alpha }{2}-k-n-\frac{1}{4},-\frac{3 \alpha}{2}-k-n+\frac{1}{4}\}, 
\end{displaymath}
while
 the nine denominator terms were
\begin{equation} 
\{-k-n+\frac{1}{2},-3 \alpha -k-n-\frac{3}{4},-3 \alpha -k-n-\frac{1}{2},-3 \alpha
   -k-n-\frac{1}{4},-\alpha -k-n+\frac{1}{2},
\end{equation}
\begin{displaymath}
 -\frac{\alpha }{2}-k-n,
-\frac{\alpha}{2}-k-n+\frac{1}{2},-\frac{k}{2}-\frac{n}{2},-\frac{k}{2}-\frac{n}{2}+\frac{1}{2}\}. 
\end{displaymath}
As the counterparts of the degree-5 two-qubit (\ref{explanatory2}) and the two-rebit (\ref{RebitExplanatory2}) results, that it had been our initial goal to 
reproduce, this new hybrid functional form yielded a 
 degree-7 two-qubit outcome
\begin{equation}
\frac{1}{512} \left(-\frac{60}{k+4}-\frac{384}{2 k+9}+\frac{15}{4
   k+9}-\frac{315}{4 k+11}+\frac{192}{4 k+17}+\frac{576}{4
   k+19}+\frac{120}{k+3}+2\right)
\end{equation}
and a degree-6 two-rebit outcome
\begin{equation}
\frac{k (k (2 k (4 k (4 k (4 k+39)+513)+2065)-4637)-12494)-7095}{2048 (k+2)^2
   (k+3) (2 k+3) (2 k+5) (4 k+13)}.
\end{equation}
\subsubsection{Separability probability estimates based on hybrid function} \label{SepProbEst}
\begin{figure}
\includegraphics[scale=.9,trim=2cm 4cm 2cm 1cm,clip]{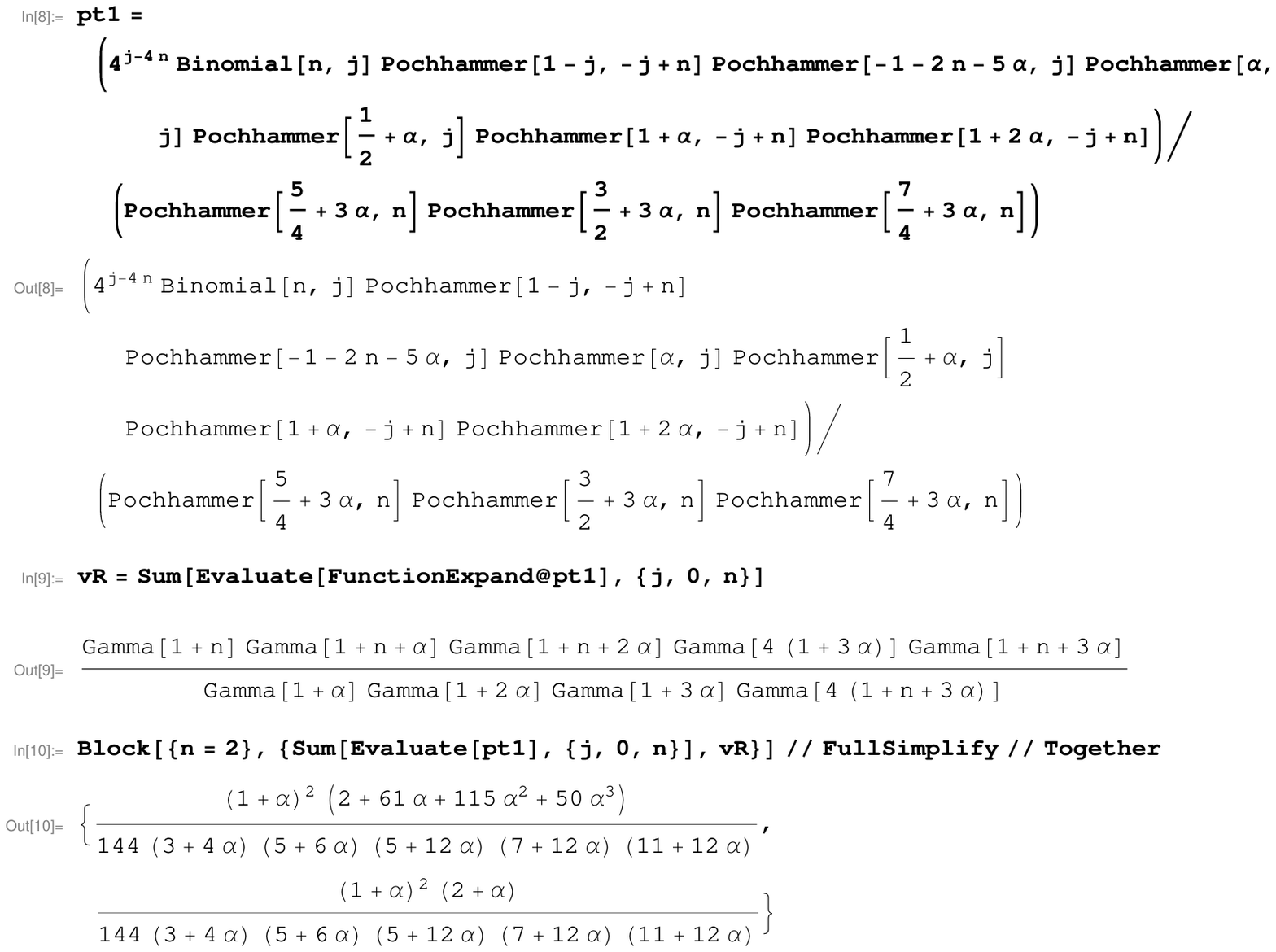}
\caption{\label{fig:Bug}Inconsistent computations in Mathematica 9.0 of 
$\langle |\rho^{PT}|^2 \rangle_{Hilbert-Schmidt}$. The first result is the {\it correct} one, given in \cite[p. 30]{MomentBased} for $k=0$.}
\end{figure}

We applied (as we have done in \cite{MomentBased,slaterJModPhys}), the Legendre-polynomial-based moment-inversion procedure of Provost \cite{Provost}, with 6,000 "hybrid moments" (cf. \cite{MaChen}) yielded by the $_{10}F_{9}$ result above. We used as a proxy for  $k=0$,
the value $k=10^{-20}$. (We have so far been unable to explicitly construct the $k=0$ limit of the hybrid $_{10}F_{9}$ function (cf. \cite[p. 30]{MomentBased}). In attempting this construction, we have uncovered--as M. Trott [Wolfram] confirmed--a certain "bug" in the use of the Mathematica command "Sum" [version 9.0] (Fig.~\ref{fig:Bug}).) 

Then, our estimate of this Hilbert-Schmidt/Bures hybrid two-rebit separability probability is 0.23631557, and in the two-qubit case, 0.079821829. We can compare this latter value with our previous "silver mean" conjecture of 
$\frac{1680 \left(\sqrt{2}-1\right)}{\pi ^8} \approx 0.0733389$ 
\cite{slaterJGP}. (Earlier still, in \cite{EarlyConjecture}, a conjecture of $\frac{8}{11 \pi^2} \approx 0.0736881$ had been advanced.) Since the evidence is highly compelling that the Hilbert-Schmidt two-qubit separability probability is
$\frac{8}{33} \approx 0.242424$ \cite{MomentBased,slaterJModPhys}, it is not implausible that this  hybrid Hilbert-Schmidt/Bures (intermediate-type) 
estimate does, in fact, exceed the conjectured value (that is, $0.079821829 > 0.0733389$), thus, providing an upper bound on the true value. 
Further, for the $\alpha =2$ two-quater(nionic)bit case, our estimate, also using $k=10^{-20}$ as a proxy for $k=0$, is 0.01245737. (We are not aware of any explicit formula that can apparently embrace the three values given above. The estimates very slowly decrease as the number of "hybrid moments" used decreases. Thus, based on the first 3,000 such "moments", the estimates were the slightly greater values, respectively, of 0.23645208, 0.079901505 and 0.012458305.)

Alternatively, if we apply the Legendre-polynomial-based moment-inversion procedure of Provost \cite{Provost} using the $_{10}F_{9}$ result, now not (nearly) nullifying $k$, but setting $k=n$, we obtain (notably more slowly converging) separability probability estimates that, concomitantly, much more rapidly decrease with the number of moments employed. Thus, for the first 3,000 such "moments", the $\alpha =\frac{1}{2}, 1$ and 2 estimates are 0.334796318, 0.2049266304 and  0.1482428189, respectively, while for 6,000 "moments" these estimates decrease rather substantially to 0.3172391218, 0.1765322036 and 0.112681718. Thus, both sets of estimates (those based on $k=0$ and those based on $k=n$) appear to provide {\it upper}, not {\it lower} bounds on the "hybrid-separability-probabilities".
\subsection{Hilbert-Schmidt measure on qubit-qutrit states}
Of course, we should also entertain the possibility
that the random-matrix-theory applicability of Dyson indices, as employed, apparently successfully, in our analogous studies based on the Hilbert-Schmidt metric 
\cite{MomentBased}, does not strictly carry over to parallel studies based on the Bures metric (cf. \cite{slaterPRA,sbz}). If that is the case, perhaps the more promising venue for progress in this general area of generating moments and associated probability distributions would be the {\it Hilbert-Schmidt} qubit-{\it qutrit} case. There, in addition to these explicit results for the two cases 
$n=1, \alpha=\frac{1}{2}, 1$  \cite[eqs. (27), (30)]{MomentBased}, 
\begin{equation} \label{rebitretrit}
\frac{\left\langle |\rho|^k |\rho^{PT}| \right\rangle_{rebit-retrit/HS}}{\left\langle |\rho|^k \right\rangle_{rebit-retrit/HS}}= 
\frac{4 k^5+40 k^4+95 k^3-220 k^2-1149 k-1170}{576 (k+4) (3 k+11) (3
   k+13) (6 k+23) (6 k+25)}
\end{equation}
and
\begin{equation} \label{qubitqutrit}
\frac{\left\langle |\rho|^k |\rho^{PT}| \right\rangle_{qubit-qutrit/HS}}{\left\langle |\rho|^k \right\rangle_{qubit-qutrit/HS}}= 
\frac{k^5+15 k^4+37 k^3-423 k^2-2558 k-3840}{72 (2 k+13) (3 k+19) (3
   k+20) (6 k+37) (6 k+41)},
\end{equation}
such as we do have available ((\ref{explanatory2}), (\ref{RebitExplanatory2})) for the analogous Bures scenarios \cite{BuresHilbert}, we additionally have the $n=2, \alpha=\frac{1}{2}$ rebit result \cite[eqs. (28), (29)]{MomentBased} (which seemed, unfortunately, too arduous to fully compute in the Bures  instance \cite{BuresHilbert} (cf. sec.~\ref{Arduous})).
If we divide the right-hand sides of (\ref{rebitretrit}) and (\ref{qubitqutrit}) by the apparently appropriate (non-hypergeometric) prefactor
\begin{equation} \label{QQprefactor}
\frac{(k+1)_n (k+\alpha +1)_n (k+2 \alpha +1)_n (k+3 \alpha +1)_n (k+4 \alpha +1)_n
   (k+5 \alpha +1)_n}{(6 k+30 \alpha +6)_{6 n}},
\end{equation}
which is based on the qubit-qutrit formula (cf. (\ref{generalformula}))
\begin{equation} \label{F2QQ}
\left\langle \left\vert \rho\right\vert ^{k}\right\rangle^{HS}_{\alpha/6 \times 6} =\frac{(1)_k (\alpha +1)_k (2 \alpha +1)_k (3 \alpha +1)_k (4 \alpha +1)_k (5
   \alpha +1)_k}{(30 \alpha +6)_{6 k}},
\end{equation}
then (\ref{rebitretrit}) is transformed to the interestingly, considerably simpler 
lower-degree form
\begin{equation}
\frac{4 k^3+20 k^2-29 k-195}{4 k^3+20 k^2+31 k+15}
\end{equation}
and (\ref{qubitqutrit}), similarly, to
\begin{equation}
\frac{k^3+7 k^2-34 k-256}{k^3+7 k^2+14 k+8}.
\end{equation}
(In the $n=2, \alpha=\frac{1}{2}$ case, we obtain the ratio of degree-8 polynomials.)

In the utility function framework, we find for the rebit-retrit case 
($\alpha=\frac{1}{2}$), that 
\begin{equation}
F_{2/6 \times 6}^{HS}(1,k,\frac{1}{2})=\frac{-10 k^3-85 k^2-235 k-210}{31104 k^5+622080 k^4+4972320 k^3+19854720
   k^2+39605856 k+31574400}
\end{equation}
\begin{displaymath}
=-\frac{5 (k+2) (k+3) (2 k+7)}{96 (k+4) (3 k+11) (3 k+13) (6 k+23) (6 k+25)},
\end{displaymath}
while for the qubit-qutrit ($\alpha=1$) case, 
\begin{equation}
F_{2/6 \times 6}^{HS}(1,k,1)=\frac{4 k^3+54 k^2+211 k+330}{3888 k^5+126360 k^4+1642140 k^3+10666890
   k^2+34633182 k+44963880}
\end{equation}
\begin{displaymath}
=\frac{k \left(4 k^2+54 k+211\right)+330}{6 (2 k+13) (3 k+19) (3 k+20) (6 k+37)
   (6 k+41)}.
\end{displaymath}
$F_{2/6 \times 6}^{HS}(1,k,\frac{1}{2})$  and $F_{2/6 \times 6}^{HS}(1,k,1)$ above are both ratios of degree-three to degree-five  polynomials, while we further found that 
$F_{2/6 \times 6}^{HS}(2,k,\frac{1}{2})$ is the ratio of a degree-ten polynomial to a degree-ten polynomial.
\subsubsection{Qubit-qutrit hybrid $_{10}F_{9}$} \label{Hyb2}
Now, we again performed the utility function type of summation indicated in (\ref{Summation}), again employing the two-qubit Hilbert-Schmidt function $F_2^{HS}(n,k,\alpha)$ given by (\ref{F2HS}), but now using for the $R$ function, not its two-qubit Bures form (\ref{RBures}), as previously in sec.~\ref{Hyb1}, but its Hilbert-Schmidt qubit-qutrit analogue, based upon (\ref{F2QQ}).
Again, the summation yielded a $_{10}F_{9}$ hypergeometric function with the anticipated 
prefactor (\ref{prefactorQQ}). Still again, as desired, the number (eight) of $\alpha$'s among the ten numerator terms was one greater than the number (seven) among the nine denominator terms. But now the argument of the $_{10}F_{9}$ function was not 1, but 
$\frac{729}{4}= \frac{3^6}{2^2}$, so it is not convergent. Additionally, the function was not exactly balanced, for the sum of the nine denominator terms 
$\{-4 \alpha -k-n,-3 \alpha -k-n-\frac{3}{4},-3 \alpha -k-n-\frac{1}{2},-3 \alpha
   -k-n-\frac{1}{4},-3 \alpha -k-n,-2 \alpha -k-n, -\alpha
   -k-n,-\frac{k}{2}-\frac{n}{2},-\frac{k}{2}-\frac{n}{2}+\frac{1}{2}\}$ minus
the sum of the ten numerator terms $\{\alpha ,\alpha +\frac{1}{2},-k,-5 \alpha -2 k-2 n-1,-5 \alpha
   -k-n-\frac{5}{6},-5 \alpha -k-n-\frac{2}{3},-5 \alpha -k-n-\frac{1}{2}, -5 \alpha
   -k-n-\frac{1}{3},-5 \alpha -k-n-\frac{1}{6},-n\}$ was not 1, but $2 + 9 \alpha$.
\section{$_{8}F_{12}$-based Bures two-qubit function} \label{8F12}
As a candidate for $F_2^{Bures}(n,k,\alpha)$ we took the ratio (using Pochhammer symbol notation) of the expression $(\mbox{num}(\alpha))_n$, given by (\ref{NumeratorFunction}), replacing $k$ in it by $k+n-1$, to the expression $128^n = 2^{7 n}$ times the product of six terms. (We note that $k=k+n-1$ for $n=1$.) These six terms were of the form $(\mbox{factor})_n$, where the six factors--$\alpha,\ldots,(6 \alpha +4 k+5)$ are those of $\mbox{den}(\alpha)$, given by (\ref{DenominatorFunction}). (Thus, for $n=1$, the candidate for $F_2^{Bures}(n,k,\alpha)$ reduces to $\frac{\mbox{num}(\alpha)}{\mbox{den}(\alpha)}$.) Then, inserting this into the principal utility function summation (\ref{Summation}), along with  $R^{Bures}(n,k,\alpha)$, 
given by (\ref{RBures}), we obtained a result that was the product of
the anticipated prefactor (\ref{BigOne2}) times an $_{8}F_{12}$ hypergeometric function with argument
$\frac{729}{32768}=\frac{3^6}{2^{15}}$. 

This function did (finally) succeed--in terms of our earlier stated objectives--in precisely reproducing our original desired two-rebit and two-qubit results (\ref{explanatory2}) and (\ref{RebitExplanatory2})--lower-degree numerator terms and all. 

The twelve denominator terms of this $_{8}F_{12}$ function were 
\begin{equation}
\{\alpha ,-k-n+\frac{1}{2},-3 \alpha -k-n-1,-\frac{3 \alpha
   }{2}-k-n-1,-\frac{3 \alpha }{2}-k-n-\frac{3}{4},-\frac{3 \alpha
   }{2}-k-n-\frac{1}{2},-\frac{3 \alpha }{2}-k-n-\frac{1}{2},
\end{equation}
\begin{displaymath}
-\frac{3 \alpha
   }{2}-k-n-\frac{1}{4},-\frac{3 \alpha }{2}-k-n,-\alpha
   -k-n+\frac{1}{2},-\frac{\alpha }{2}-k-n,-\frac{\alpha
   }{2}-k-n+\frac{1}{2}\}
\end{displaymath}
and seven of the numerator terms were 
\begin{equation}
\{-2 \alpha -\frac{4 k}{3}-\frac{4 n}{3}-\frac{4}{3},-2 \alpha -\frac{4
   k}{3}-\frac{4 n}{3}-1,-2 \alpha -\frac{4 k}{3}-\frac{4 n}{3}-\frac{2}{3},-2
   \alpha -\frac{4 k}{3}-\frac{4 n}{3}-\frac{2}{3},
\end{equation}
\begin{displaymath}
-2 \alpha -\frac{4
   k}{3}-\frac{4 n}{3}-\frac{1}{3},-2 \alpha -\frac{4 k}{3}-\frac{4
   n}{3},-n\}.
\end{displaymath}
(We note that six of the eight numerator terms contain  $-2 \alpha -\frac{4 k}{3} -\frac{4 n}{3}$, with a seventh being $-n$, and also that all but one of the denominator terms contain $-k-n$.) The remaining (eighth) numerator term was a polynomial quartic (fourth-degree) in $\alpha,k$ and $n$ with thirty summands, that is, $\{-1280 \alpha ^4-128 \alpha ^3+908 \alpha ^2-460 \alpha -160 \alpha 
   k^3+64 k^3-960 \alpha ^2 k^2+224 \alpha  k^2-480 \alpha  k^2 n+192 k^2 n+64
   k^2-1920 \alpha ^3 k+128 \alpha ^2 k+582 \alpha  k-480 \alpha  k n^2+192 k
   n^2-1920 \alpha ^2 k n+448 \alpha  k n+128 k n-152 k-160 \alpha  n^3+64
   n^3-960 \alpha ^2 n^2+224 \alpha  n^2+64 n^2-1920 \alpha ^3 n+128 \alpha ^2
   n+582 \alpha  n-152 n+75\}$. 

However, we were compelled to conclude that this result could be by no means the definitive solution for which we had been searching, as it  led, apparently, to implausible ($>1$) probabilistic results upon use of the moment-inversion procedure \cite{Provost}. Further, in addition to not being balanced, not having argument 1, and lacking a numerator term of $-k$, it had eleven denominator terms containing $\alpha$, and only seven such numerator terms (thus, not meeting the asymptotic criterion suggested by Dunkl). Also, obviously, it is not of the form $_{p}F_{p-1}$.
\section{Computations for  Bures $n=2, \alpha=\frac{1}{2}$ case} \label{Arduous}
To obtain $\left\langle \left\vert \rho^{PT}\right\vert^n \left\vert \rho\right\vert
^{k}\right\rangle^{Bures}_{2-rebits}$, for $n=2$, we need to compute the average with respect to  the corresponding Bures normalized measure \cite[eq. (3.19)]{szBures} \cite[eq. (42)]{BuresHilbert}
\begin{equation} \label{Buresformula1}
P^{Bures}(\alpha) =\frac{128}{\pi} \frac{\Pi_{i<j}^{4} \Big(\frac{{|\lambda_i -\lambda_j|}^2}{\lambda_i+\lambda_j}\Big)^{\alpha}}{\sqrt{\Pi_{i=1}^{4} \lambda_i}}
\end{equation}
with $\alpha=\frac{1}{2}$, of 
the product of
\begin{equation}
\left(\lambda _1 \lambda _2 \lambda _3 \lambda _4\right){}^k
\end{equation}
and
\begin{equation} \label{lotsofmonomials}
\frac{\lambda _4^8}{576}+\frac{1}{252} \lambda _3 \lambda _4^7-\frac{103 \lambda
   _3^2 \lambda _4^6}{8400}-\frac{89 \lambda _2 \lambda _3 \lambda
   _4^6}{6300}-\frac{197 \lambda _3^3 \lambda _4^5}{14700}-\frac{101 \lambda _2
   \lambda _3^2 \lambda _4^5}{2450}-\frac{2981 \lambda _1 \lambda _2 \lambda _3
   \lambda _4^5}{44100}
\end{equation}
\begin{displaymath}
+\frac{43091 \lambda _3^4 \lambda
   _4^4}{2116800}+\frac{289 \lambda _2 \lambda _3^3 \lambda
   _4^4}{9450}+\frac{143 \lambda _2^2 \lambda _3^2 \lambda
   _4^4}{3920}-\frac{5641 \lambda _1 \lambda _2 \lambda _3^2 \lambda
   _4^4}{44100} +\frac{433 \lambda _2^2 \lambda _3^3 \lambda
   _4^3}{22050}
\end{displaymath}
\begin{displaymath}
+\frac{28181 \lambda _1 \lambda _2 \lambda _3^3 \lambda
   _4^3}{132300}+\frac{1091 \lambda _1 \lambda _2^2 \lambda _3^2 \lambda
   _4^3}{2450}+\frac{59441 \lambda _1^2 \lambda _2^2 \lambda _3^2 \lambda
   _4^2}{117600},
\end{displaymath}
where the $\lambda$'s denote the eigenvalues of the $4 \times 4$ density matrix $\rho$ 
(cf. \cite[eq. (20)]{BuresHilbert}).

It was reported in \cite[eq. (32)]{BuresHilbert} that
\begin{equation} \label{biggy2}
\left\langle \left\vert \rho^{PT}\right\vert^2 \right\rangle_{2-rebits}^{Bures} =  \frac{50654227}{1307993702400} =
\frac{13 \times 101 \times 173 \times 223}{2^{23} \times 3^4 \times 5^2 \times 7 \times 11} \approx 0.0000387267.
\end{equation}
Also, the expected values of certain--but not all--of the eigenvalue monomials employed in (\ref{lotsofmonomials}) were presented there, as well \cite[eqs. (28)-(31), (33)-(36)]{BuresHilbert}. The three eigenvalue monomials of the fifteen employed in (\ref{lotsofmonomials}) for which we still lack expected value formulas 
in this $n=2$ two-rebit case are $\lambda_4^8, \lambda_3 \lambda_4^7$ and $\lambda_3^2 \lambda_4^6$.
With this available knowledge, we are able to obtain
\begin{equation} \label{almost}
\left\langle \left\vert \rho^{PT}\right\vert^2 \left\vert \rho\right\vert
^{k}\right\rangle^{Bures}_{2-rebits} = \left\langle \left(\frac{\lambda _4^8}{576}+\frac{1}{252} \lambda _3 \lambda _4^7-\frac{103
   \lambda _3^2 \lambda _4^6}{8400}\right) \left(\lambda _1 \lambda _2 \lambda
   _3 \lambda _4\right){}^k \right\rangle +
\end{equation}
\begin{displaymath}
-\frac{4^{-4 k-13}  \left(-\frac{1}{4}\right)_{k+1}
   \left(\frac{1}{4}\right)_{k+1}S }{3472875 (k+2) (k+3) (2 k+1) (2 k+3) (2 k+5)
   (3)_{k+1} \left(\frac{11}{2}\right)_{k+1}},
\end{displaymath}
where $3472875 = 3^4 \cdot 5^3 \cdot 7^3$ and 
\begin{equation}
S=2909089792 k^9+58511409152 k^8+464093464048 k^7+1738045720352 k^6+2166731989792
   k^5
\end{equation}
\begin{displaymath}
-6313467831760 k^4-29574963265176 k^3-49276400150880 k^2-38483212259637
   k-11376862535850.
\end{displaymath}
\subsection{$k=0$}
For $k=0$, we have (\ref{biggy2}) that
\begin{equation}
\left\langle \left\vert \rho^{PT}\right\vert^2 \right\rangle_{2-rebits}^{Bures} =  \frac{50654227}{1307993702400} = \left\langle \left(\frac{\lambda _4^8}{576}+\frac{1}{252} \lambda _3 \lambda _4^7-\frac{103
   \lambda _3^2 \lambda _4^6}{8400}\right) \right\rangle
-\frac{10835107177}{5273830608076800},
\end{equation}
so that 
\begin{equation} \label{start}
\left\langle \left(\frac{\lambda _4^8}{576}+\frac{1}{252} \lambda _3 \lambda _4^7-\frac{103
   \lambda _3^2 \lambda _4^6}{8400}\right) \right\rangle = \frac{215072950441}{5273830608076800} =\frac{31 \cdot 53 \cdot 130902587}{2^{29} \cdot 3^6 \cdot 5^2 \cdot 7^2 \cdot 11} \approx 0.000040781163.
\end{equation}
If we can extend this line of computation from the $k=0$ case above to $k =1, 2, 3,\ldots,N$, for $N$ sufficiently large, then we should be able to apply the 
FindSequenceFunction of Mathematica to determine the underlying formula, and thus, via (\ref{almost}) obtain a general expression for $\left\langle \left\vert \rho^{PT}\right\vert^2 \left\vert \rho\right\vert
^{k}\right\rangle^{Bures}_{2-rebits}$. (A numerical integration procedure yielded 
0.000040781133 for (\ref{start}) and $1.420551358 \times 10^{-9}, 3.522724342 \times 10^{-13}, 1.8925708934 \times 10^{-16}$ and $1.54994576705 \times 10^{-19}$ for the succeeding cases $k=1, 2, 3, 4$.) 

With our results (\ref{start}) for the case $k=0$, we were able to determine that
\begin{equation}
\left\langle \lambda _4^8 \right\rangle = \frac{29607386147749}{1318457652019200} =\frac{29607386147749}{2^{27} \cdot 3^6 \cdot 5^2 \cdot 7^2 \cdot 11},
\end{equation}
\begin{displaymath}
\left\langle \lambda_3 \lambda _4^7  \right\rangle = \frac{10306738882511}{5273830608076800} =\frac{677 \cdot 29927 \cdot 508709}{2^{29} \cdot 3^6 \cdot 5^2 \cdot 7^2 \cdot 11}
\end{displaymath}
and 
\begin{displaymath}
\left\langle  \lambda_3^2 \lambda _4^6  \right\rangle = \frac{102540856051}{210953224323072} = \frac{2111 \cdot 48574541}{2^{29} \cdot 3^6 \cdot 5^2 \cdot 7^2 \cdot 11}.
\end{displaymath}
\subsection{$k=1$}
Further computations appear to indicate that for $k=1$,
\begin{equation} \label{rebitn=2}
\left\langle \left\vert \rho^{PT}\right\vert^2 \left\vert \rho\right\vert^k
\right\rangle^{Bures}_{2-rebits} = \frac{11395427}{9630347469783040} =
\frac{67 \cdot 170081}{2^{38} \cdot 5 \cdot 7^2 \cdot 11 \cdot 13} \approx 1.183283058 \cdot 10^{-9}.
\end{equation}

We are presently pursuing further such computations ($k>1$), as well as laying groundwork for the $n =2, \alpha =1$ case \cite[p. 11]{BuresHilbert}. (We note that if 
(\ref{almost}) and (\ref{rebitn=2}) are divided by $\left\langle \left\vert \rho\right\vert ^{k}\right\rangle^{HS}_{\frac{1}{2}}$--given by (\ref{generalformula})--then they  more directly correspond 
with the structure of the $n=1$ moment formulas (\ref{explanatory2}) and (\ref{RebitExplanatory2})).
\subsection{Completed computation}
We were ultimately able to obtain for this $n =2, \alpha =\frac{1}{2}$ scenario that 
$\frac{\left\langle \left\vert \rho^{PT}\right\vert^2 \left\vert \rho\right\vert
^{k}\right\rangle^{Bures}_{2-rebits}}{\left\langle \left\vert \rho\right\vert ^{k}\right\rangle^{Bures}_{2-rebits}}$ is the ratio of two polynomials in $k$ of degree-10.
The denominator of this polynomial is (cf. (\ref{RebitExplanatory2})
\begin{equation}
2^{12} (k+2)^2 (k+3)^2 (2 k+3) (2 k+5)^2 (2 k+7) (2 k+9) (2 k+11),
\end{equation} 
while the numerator is
\begin{equation}
4096 k^{10}+86016 k^9+730624 k^8+3191808 k^7+7842576 k^6 +
\end{equation} 
\begin{displaymath}
16125680 k^5+63736088
   k^4+248378840 k^3+557112761 k^2+644925323 k+303925362.
\end{displaymath}
This numerator is expressible as
\begin{equation}
k (k (8 k (k (2 k (k (32 k (k (8 k
   (k+21)+1427)+6234)+490161)+1007855)+7967011)+31047355)+557112761)
\end{equation}
\begin{displaymath}
+644925323)+
   303925362.
\end{displaymath}

At this point, we were able to compute the utility function $F_2^{Bures}(2,k,\frac{1}{2})$, as the ratio of degree-10 polynomials. Together with our earlier $n=1$ results ((\ref{F2Complex}) and (\ref{F2Real}))--both ratios of degree-3 polynomials to degree-5 polynomials--for $F_2^{Bures}(1,k,1)$ and $F_2^{Bures}(1,k,\frac{1}{2})$, it did not appear to us that it would be possible to construct
a more general explanatory formula for $F_2^{Bures}(n,k,\alpha)$, based on these three results, with a form strictly parallel to that (\ref{F2HS}) (that is, the ratios of products of Pochhammer symbols) adhered to by $F_2^{HS}(n,k,\alpha)$.
\section{Concluding Remarks}
In the generalized two-qubit Hilbert-Schmidt analysis \cite{MomentBased}, the non-hypergeometric factor employed (\ref{BigOne}) is an obvious deduction of the moment formula (\ref{generalformula}). In this study, we have assumed such a pattern to also hold in the Bures two-qubit  case and Hilbert-Schmidt qubit-qutrit case. This assumption needs to be more critically examined--but seems difficult to do so in the absence of more extensive moment exact, symbolic calculations than are presently available. It further behooves us to ascertain whether the hybrid Hilbert-Schmidt/Bures $_{10}F_{9}$ result (sec.~\ref{Hyb1}) 
fulfills the (Hausdorff/Hankel-matrix) requirements for the moments of an actual probability distribution
(over $|\rho^{PT}| \in [-\frac{1}{16},\frac{1}{256}]$) \cite{HolgerStudden}. (We note that, typically, such requirements are formulated for probability distributions defined over the unit interval, while, in the case at hand, $|\rho^{PT}| \in [-\frac{1}{16},\frac{1}{256}]$, so certain transformations would be required.) 
The 
$_{10}F_{9}$ result leads us to speculate that possibly a ("pure"-nonhybrid) 
Bures two-qubit function may also be of such a $_{10}F_{9}$ form.

Mathematica analyses have convinced us that if either the generalized two-qubit Bures or  qubit-qutrit Hilbert-Schmidt moments--adhering to the two-qubit Hilbert-Schmidt hypergeometric paradigm (\ref{BigOne}) established in \cite{MomentBased}--takes the form of the product of the associated prefactor (see (\ref{BigOne2}), (\ref{QQprefactor})) times a $_{p}F_{p-1}$ hypergeometric function containing $-n$ and $-k$ among its numerator terms, then it would be necessary that $p>9$. We continue to explore such a possibility.

\begin{acknowledgments}
I would like to express appreciation to the Kavli Institute for Theoretical
Physics (KITP) for computational support in this research. C. Dunkl supplied several 
useful insights.
\end{acknowledgments}

\bibliography{JointAnalyses}

\end{document}